\documentclass[%
 aip, rsi, reprint, amsmath, amssymb]{revtex4-2}

\usepackage[separate-uncertainty=true, multi-part-units=single]{siunitx} % Good SI unit and number formatting
\usepackage{braket}		% Dirac's bra-ket notation
\usepackage{graphicx}% Include figure files
\usepackage{dcolumn}% Align table columns on decimal point
\usepackage{bm}% bold math
\usepackage{multirow}
\usepackage[utf8]{inputenc}
\usepackage[T1]{fontenc}
\usepackage{mathptmx}
\usepackage{etoolbox}
\usepackage{bbold}

\makeatletter
\def\@email#1#2{%
 \endgroup
 \patchcmd{\titleblock@produce}
  {\frontmatter@RRAPformat}
  {\frontmatter@RRAPformat{\produce@RRAP{*#1\href{mailto:#2}{#2}}}\frontmatter@RRAPformat}
  {}{}
}%
\makeatother
\begin{document}

\title[Characterization of optical aberrations with scanning pentaprism for large collimators]{Characterization of optical aberrations with scanning pentaprism for large collimators}
% Force line breaks with \\
\author{Youn Seok Lee}
\affiliation{Institute for Quantum Computing and Department of Physics and Astronomy, University of Waterloo, Waterloo, Ontario N2L 3G1, Canada}
  \email{ys25lee@uwaterloo.ca}
  
\author{Kimia Mohammadi}
 \affiliation{Institute for Quantum Computing and Department of Physics and Astronomy, University of Waterloo, Waterloo, Ontario N2L 3G1, Canada}

\author{Thomas Jennewein}
\affiliation{Institute for Quantum Computing and Department of Physics and Astronomy, University of Waterloo, Waterloo, Ontario N2L 3G1, Canada}

\date{\today}% It is always \today, today,

\begin{abstract}
We present a practical apparatus for characterizing optical aberrations of large collimation mirrors and lenses, and give a detailed analysis of wavefront-detection errors. We utilize a scanning pentaprism technique for precise measurements of local wavefront slopes, and reconstruct transmitted wavefronts via a conventional least-squares method. Our proof-of-principle experiment demonstrates transverse linear measurements of transmitted wavefronts for a \SI{20.3}{\centi\meter} lens developed for Canada's Quantum Encryption and Science Satellite (QEYSSat) mission. Our demonstration shows the wavefront-detection precision better than $0.01\lambda$ and the divergence-angle resolution less than \SI{20}{\micro\radian} over the range of \SI{40}{\centi\meter}. We model our optical setup using three-dimensional raytracing and find good quantitative agreement between experimental results and theoretical predictions which validates our methodology.
\end{abstract}

\maketitle

\section{\label{sec:intro}Introduction}

Telescopes for launching (near) collimated light with the aperture diameter greater than \SI{20}{\centi\meter} have been used in diverse research fields including laser ranging satellites~\cite{Wilkinson2019}, laser guide stars~\cite{Parenti:94}, lidar~\cite{Jaboyedoff2012}, and free-space optical communications~\cite{Kaushal2017,Ursin2007}. All these applications require laser fields that are carefully collimated and have very little optical aberrations. Therefore, it is a crucial task to precisely measure and characterize the optical quality and alignment of systems.

Commercially available devices for characterizing optical aberrations such as shearing interferometers limit aperture sizes of test optics typically up to \SI{10}{\centi\meter}, and larger optics require additional beam expanders. Also, the measurement setups are usually heavy, and not portable or customized for various types of optics. Thus, several custom-built measurement schemes have been reported for testing large refractive and reflective collimators. One is to laterally scan a Shack-Hartmann sensor across the aperture of test optics~\cite{Kiikka2006}. This method requires an additional calibration and/or complex image post-processing. One other method uses a Shack-Hartmann sensor attached to the eyepiece of telescopes and star light as a point source with minimal turbulence effect via averaging images over a long measurement time~\cite{Potanin2009,Drabek2017}.

Over the past few decades, a scanning pentaprism technique, which was originally proposed for collimation tests of telescopes~\cite{Wilson1990}, has been applied for precise topographic measurements for both curved~\cite{Su2008,Allen2010} and flat meter-sized mirrors~\cite{Yellowhair2007,Qi2019} as well as \SI{500}{\milli\meter} wafers~\cite{Geckeler2002}. It allows absolute measurements of local angle of incidences, or wavefront slopes, across a transverse field profile. With optimal usage of pentaprism to minimize its motion-induced errors, sub-nano meter precision of measuring surface roughness can readily be achieved~\cite{Geckeler2006}. However, most previous demonstrations have been focused on measuring topography of meter-sized reflective optics.

In this report, we present a practical and cost-effective apparatus with a scanning pentaprism to characterize optical aberrations of collimated light. We demonstrates transverse linear measurements of the transmitted wavefront for a \SI{20.3}{\centi\meter} lens at different positions and tilt angles. We develop a portable measurement system which consists of only commercial off-the-shelf items, yet performs the precision better than $0.01\lambda$ up to \SI{40}{\centi\meter} aperture size of the test optics.

\begin{figure*}[t]
	\centering
	\includegraphics[width=0.8\linewidth]{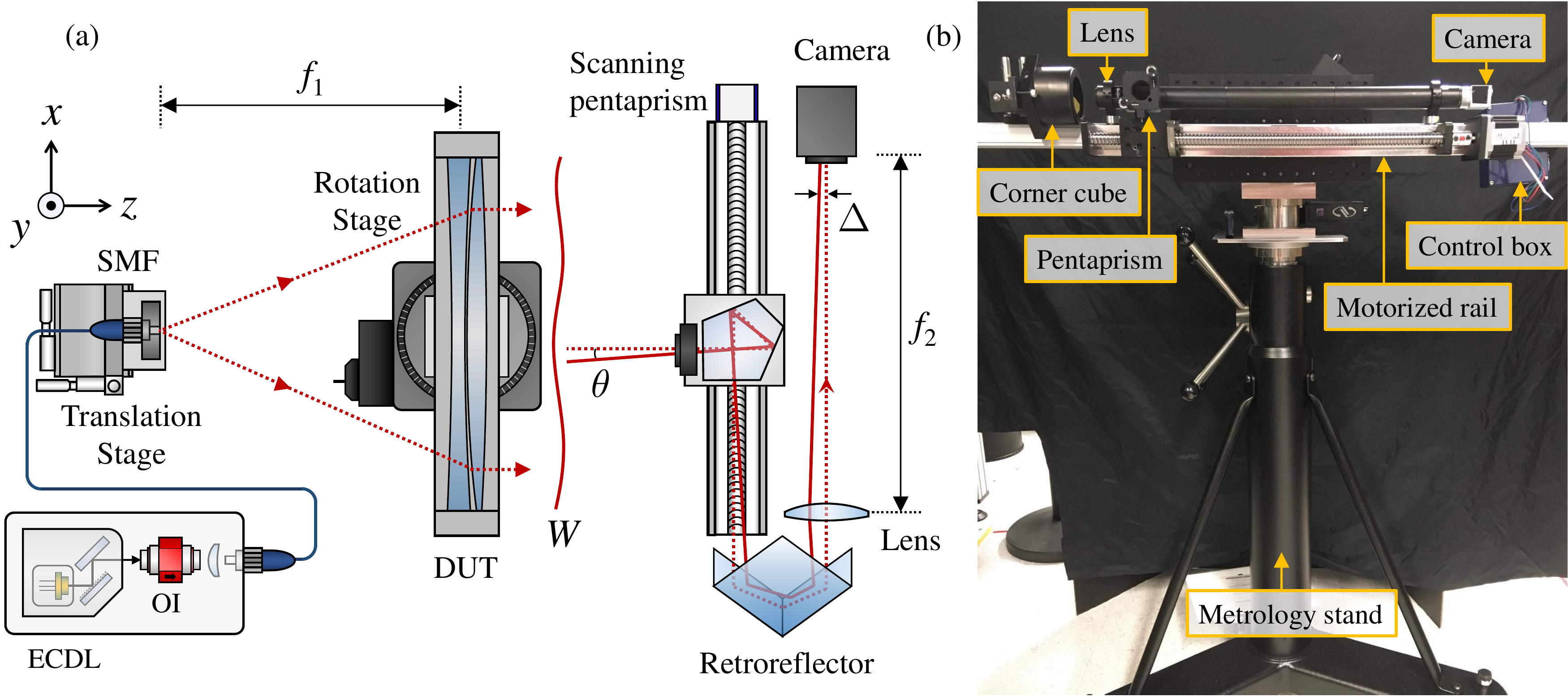}
	\caption{(a) A schematic diagram of the experimental setup; ECDL, external cavity diode laser; SMF, single-mode fiber; OI, optical isolator; DUT, device under test. A pentaprism deflects the incident light by \SI{90}{\degree} and a convex lens converts the incident angle to the position of centroid spot on the camera. A motorized linear stage moves a pentaprism across the diameter of the incident light. (b) A photo of the scanning pentaprism system.
	}
	\label{fig:ExpSetup}
\end{figure*}

\section{\label{sec:design}System design and methodology}

Figure~\ref{fig:ExpSetup}(a) shows our experimental setup for characterizing wavefront distortions of a device under test (DUT). Our DUT is a \SI{20.3}{\centi\meter} achromatic doublet ($f_{1}=$\SI{243.8}{\centi\meter}) designed and manufactured for our quantum optical transmitter for Canada's Quantum Encryption and Science Satellite (QEYSSat) mission~\cite{Jennewein_qeyssat2018,Kimia2021}. A continuous-wave laser operating at the wavelength of $\lambda$=\SI{785}{\nano\meter} is coupled to a single-mode fiber which is mounted on a motorized translation stage. The DUT is installed on a rotation stage (RVS80CC, Newport) enabling the tilt around the vertical y-axis with accuracy better than \SI{0.01}{\degree}. We determine the focal position by back-propagating a collimated visible laser through the lens. This allows precisely positioning the launch fiber in the xy-plane and a rough estimation of the focal plane in the z-direction due to long Rayleigh length of the DUT. The fiber produces diverging light propagating through the DUT and the transmitted light is then roughly collimated.

The measurement apparatus consists of a pentaprism (CCM1-PS932, Thorlabs) and a pinhole that are mounted on a motorized linear stage (FSL40, FUYU) actuated by a stepper motor over the travel range of \SI{40}{\centi\meter}. The guiding rail is straight with deviation no more than \SI{0.085}{\milli\meter}. The linear stage moves the pentaprism across the aperture of the DUT and samples the light through the pinhole with a diameter of \SI{4}{\milli\meter}. At each position of the pentaprism, the sampled light with the angle of incidence $\theta$ is deflected by \SI{90}{\degree}. Then, an imaging lens ($f_{2}=$\SI{500}{\milli\meter}) converts the angle $\theta$ to the position of the focused spot $\Delta=f_{2}\tan(\theta)$. In our setup, the pentaprism is the only moving part during the measurement. The beauty of using the pentaprism is that the centroid in the xz-plane is not sensitive to three pentaprism's rotation angles to the first order. Also, note that the spot position is insensitive to translational shift of the incident light to the first order. Therefore, the designed scheme is capable of detecting the variation of the incident angle caused only by the thickness irregularity of the test optic. The CMOS camera (acA1920-40um, Basler) records the variation $\Delta$ with subpixel centroid algorithms~\cite{subpixelAccuracy}. Under the paraxial approximation for the imaging lens, the relationships between the local wavefront slope $S$, the transmitted wavefront $W(x,y)$, the variation of the centroid $\Delta$, and the incident angle $\theta$ may be written as 
\begin{equation}
S=\frac{\partial W(x,y)}{\partial x}=\tan(\theta)=\frac{\Delta}{f_{2}}.
\end{equation}

In our proof-of-principle demonstration, we perform a transverse linear measurement of the transmitted wavefront. A simple numerical integration of the measured slopes over the scanning range yields the reconstruction of the transmitted wavefront: $S_{i}=(W_{i+1}-W_{i})/h$, where $h$ is the distance between the adjacent pentaprism positions. The numerical relation between the measured slope values $S_{i}$ and the reconstructed wavefront $W_{i}$ can be expressed by a $N$ by $N+1$ sparse matrix $\mathbf{A}$ as
\begin{equation}
\vec{S} = \mathbf{A}\vec{W},
\end{equation}
where $N$ is the number of the pentaprism positions during the scanning. To solve this linear equation, we first multiply the transpose of the matrix $\mathbf{A}$ to both sides of the equation: $\mathbf{A}^{\text{T}}\vec{S} = \mathbf{A}^{\text{T}}\mathbf{A}\vec{W}$. Since the matrix $\mathbf{A}^{\text{T}}\mathbf{A}$ is singular, we impose an additional condition of the zero-mean value of the wavefront $\vec{W}$ across the test optic, as discussed in~\cite{Southwell1980}. We add an additional row of ones to the matrix $\mathbf{A}$ to construct the $N+1$ by $N+1$ extended matrix $\mathbf{A}_{e}$. Then, the solution of the transmitted wavefront $\vec{W}$ can be obtained by multiplying the inverse matrix of $\mathbf{A}_{e}^{\text{T}}\mathbf{A}_{e}$ as
\begin{equation}
\vec{W} = (\mathbf{A}_{e}^{\text{T}}\mathbf{A}_{e})^{-1}\mathbf{A}^{\text{T}}\vec{S}.
\end{equation}
The uncertainty of the wavefront measurement $\delta W$ was simply estimated from the standard deviation of the wavefront-slope measurements multiplied by the distance between two adjacent pentaprism position: $\delta W = \delta S \times h$. This is valid because the measured slope values already include the errors of positioning the pentaprism.

Our measurement setup is built on a \SI{34}{\milli\meter}$\times$\SI{34}{\milli\meter}$\times$\SI{1000}{\milli\meter} aluminum extrusion rail (XT34-1000, Thorlabs), as shown in Figure~\ref{fig:ExpSetup}(b). To reduce the total length of the apparatus while keeping a long focal length of the imaging lens, we used a corner cube reflector to fold the sampled light. The scanning pentaprism system is mounted on a 2-axis rotational stage for alignment of the system as well as for scanning in different directions across the test optic. The whole assembly is mounted on a portable and height-adjustable heavy duty metrology stand (233 Series, BRUNSON), thereby constructing a mobile wavefront sensor for large optics. 

\begin{figure*}[t]
	\centering
	\includegraphics[width=0.7\linewidth]{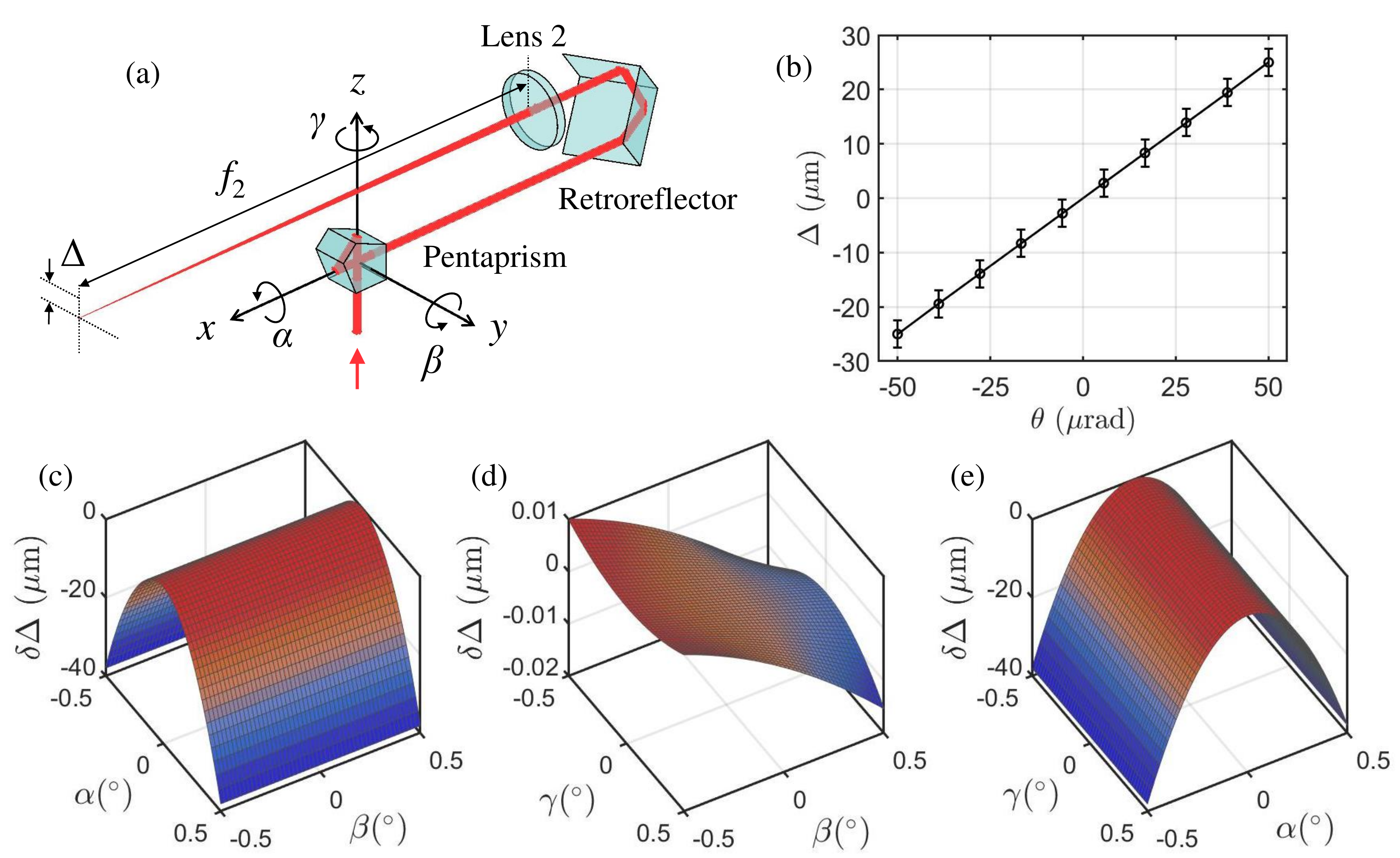}
	\caption{(a) Optical configuration of our wavefront measurement apparatus. (b) Monte-Carlo estimation of the precision of the wavefront slope measurement $\Delta$ with randomly distributed rotation angles of the pentaprism. (c)--(e) The centroid variation $\delta\Delta$ as a function of the rotation angles of the pentaprism.
	}
	\label{fig:DesignAnalysis}
\end{figure*}

\section{\label{sec:analysis}System analysis}

According to our design specifications, the optical path difference of our DUT is estimated to be 0.066$\lambda$ of Peak-to-Valley (PV) value which is mainly caused by spherical aberration. However, the manufacturing process of the lens causes additional aberrations from the surface irregularity of about $W_{\text{PV}}=0.17\lambda$. Our goal is the resolution of the divergence-angle measurement to be \SI{2.5}{\micro\radian} which yields the wavefront-measurement precision of \SI{5}{\nano\meter} for a \SI{2}{\milli\meter} step size of scanning pentaprism. Under the paraxial approximation of the imaging lens ($f_{2}=$\SI{500}{\milli\meter}), the \SI{4}{\milli\meter} aperture size of the iris yields a spot size of \SI{240}{\micro\meter} which covers more than 40 pixels of our imaging sensor. With a conservative assumption of the centroid estimation precision being 1/10 of the pixel size of \SI{5.86}{\micro\meter}~\cite{subpixelAccuracy}, the measurement precision of the divergence angle $\theta$ is estimated to be approximately \SI{1}{\micro\radian} in the absence of any systematic errors. Also, our imaging sensor size (\SI{11.3}{\milli\meter} $\times$ \SI{7.1}{\milli\meter}) gives the dynamic range of our wavefront sensor of the PV value of greater than 50$\lambda$.

To estimate the measurement precision of centroid positions after accounting higher order effects, we investigated the variation of centroid positions $\delta\Delta$ as a function of the rotation angles of the pentaprism via three-dimensional raytracing, as shown in Figure~\ref{fig:DesignAnalysis}. In our model, the rotations were performed in the order z-, y- and x-axis (labeled $\gamma$, $\beta$, $\alpha$, respectively) and we assume the perfect pentaprism: no tilt-angle between adjacent surfaces. The Figure~\ref{fig:DesignAnalysis}(c)--(e) shows the variation of the spot position $\delta\Delta$ for normal incident light $\theta=0$\si{\degree} under pentaprism rotation from \SIrange{-0.5}{0.5}{\degree}. Rotation around the x-axis ($\alpha$ rotation) showed the most significant impact on the slope measurement, with quadratic response. Meanwhile, the centroid measurement was relatively insensitive to the $\beta$ and $\gamma$ rotations. The $\alpha$ rotations (around the x-axis) during the measurement can be suppressed by monitoring the retro-reflected light with auto-correlators and feedback, as demonstrated in other deflectometries~\cite{Yellowhair2007,Qi2019}. It is worth noting that, although the variation is on the order of tens of nanometers, the centroid depends on the $\beta$ rotation (around the y-axis), which is attributed to the spherical aberration of the imaging lens.

We performed a Monte-Carlo analysis to determine the precision of the slope measurement. We varied the incident angle $\theta$ from \SIrange{-50}{50}{\micro\radian} with \SI{11.1}{\micro\radian} increments and sampled ten thousands randomly distributed pentaprism-rotation angles in a range from \SIrange{-0.25}{0.25}{\radian} at each incident angle. The mean value and the standard deviation of the centroids $\Delta$ were calculated as a function of the angles of incidence, as shown in Figure~\ref{fig:DesignAnalysis}(b). The centroid position uncertainty characterized by the averaged standard deviations was estimated to be \SI{2.5}{\micro\meter}, which is translated to the divergence-angle uncertainty of \SI{5}{\micro\radian}.

\section{\label{sec:result}Results}

\begin{figure*}[t]
	\centering
	\includegraphics[width=0.8\linewidth]{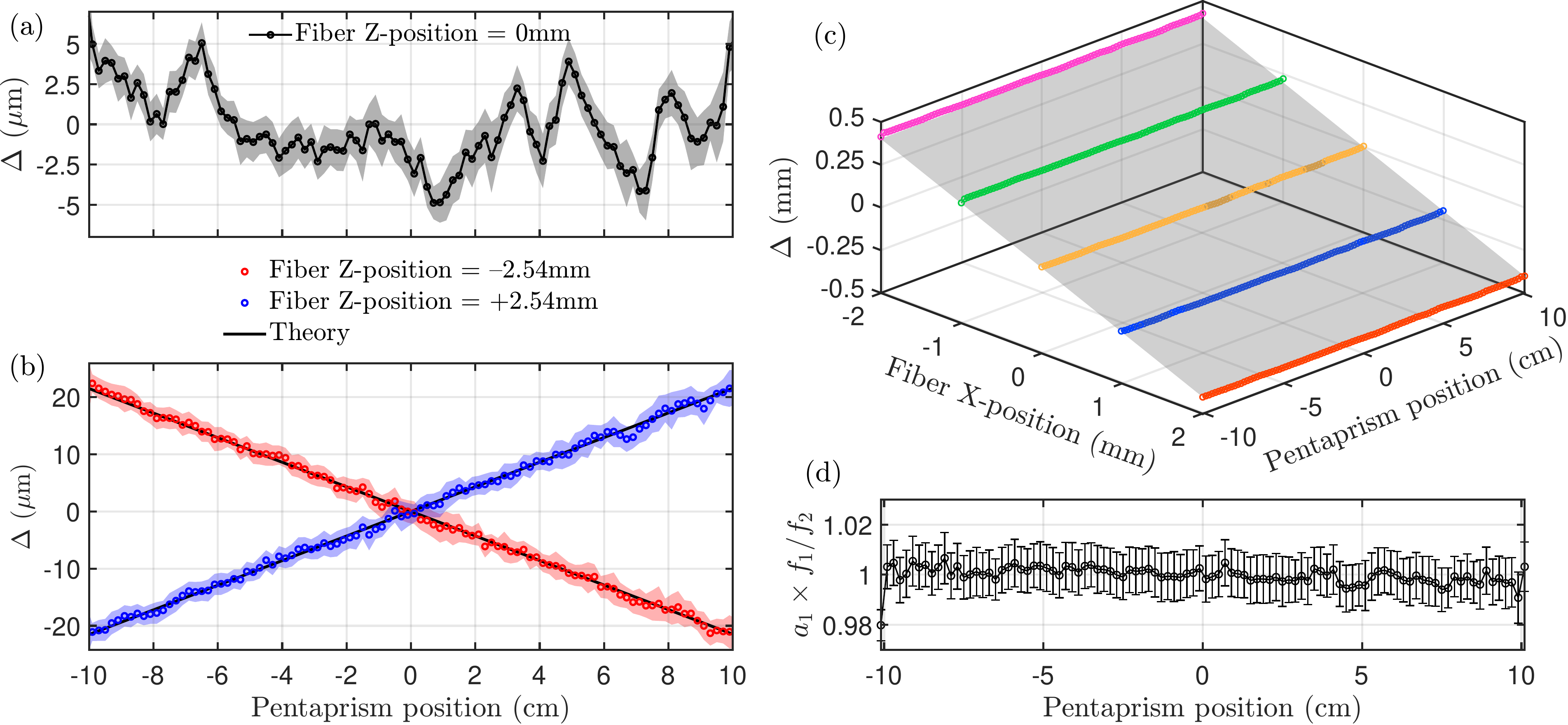}
	\caption{(a) The measured centroid position $\Delta$ values as a function of the pentaprism position for collimated light and (b) for de-focused light. Black solid lines represent the theoretical prediction calculated from a three-dimensional raytracing. (c) Characterization of the centroid position measurement with the scanning pentaprism. At each pentaprism position, the centroid position is measured for five different positions of the optical fiber in the x-direction. The dependency of the measured five centroid position values to the corresponding fiber positions are linearly fit to obtain the focal ratio of the DUT to the imaging lens ($f_{1}/f_{2}$). (d) The slope coefficient $a_{1}$ of the linear function $y=a_{1}x+a_{b}$ multiplied by $f_{2}/f_{1}$ is plotted as a function of the pentaprism position.
	}
	\label{fig:Result1}
\end{figure*}

To experimentally obtain the wavefront-measurement precision including all systematic errors, we performed the characterization with sufficient redundency. The \SI{20}{\centi\meter} measurement range of the scanning pentaprism is discretized with step size of \SI{2}{\milli\meter}. At each position, the camera adjusts its exposure time to keep a good signal-to-noise ratio (SNR$>$10) while ensuring no saturated pixels. Then, we captured twenty frames of images and calculated the centroid positions in the x-direction for all images. The full scan of the pentaprism was repeated five times, and therefore at each pentaprism position we collected one hundred centroid values. Total runtime of our experiment is about 30 minutes.

\subsection{Wavefront slope measurement}

We measured the centroid positions at various positions of the launch fiber to characterize and calibrate our measurement apparatus. Two parameters must be experimentally obtained: the focal lengths of the DUT ($f_{1}$) and the imaging lens ($f_{2}$). First, we adjusted the fiber position in the z-direction to find the minimal divergence angle, as shown in Figure~\ref{fig:Result1}(a). We observed that the variation of centroid values was kept within a PV value of $\Delta_{\text{PV}}=$\SI{9.53}{\micro\meter}. The focal length of the DUT $f_{1}=$\SI{243(1)}{\milli\meter} was measured with a laser-distance measurer, which showed good agreement with the design parameter. The centroid variation $\Delta$ is mainly attributed to the thickness variation of the DUT. The averaged standard deviation of the centroid measurements was found to be \SI{0.24}{\micro\meter} which is twenty times smaller than a pixel size of the camera being used. Note that the presented results were averaged over five times of a full scan of the measurement. This excellent repeatibility shows a great stability of our measurement apparatus.

To find the focal length of the imaging lens, we shifted the launch fiber in the z-direction by $\pm$\SI{2.54}{\milli\meter} and obtained the difference of centroid variations by subtracting the measured values at the original fiber position, as shown in Figure~\ref{fig:Result1}(b). This removes the wavefront error caused by the DUT's surface irregularity, and leaves contributions mainly from the beam divergence or convergence owing to the de-focused fiber position. Thus, the obtained results can be accurately predicted via three-dimensional raytracing. We used the measured focal length of the test optic and compared the experimental results with the predicted values (black solid lines) to determine the focal length of the imaging lens $f_{2}=$\SI{500(1)}{\milli\meter}. We characterized the closeness between the theory and experiment by a statistical parameter $R^{2}=1-\sum_{i=1}^{n}(y_{i}-f(x_{i}))^2/\sum_{i=1}^{n}(y_{i}-\bar{y})^2$ with $\bar{y}$ denoting the mean value of $y_{i}$. Here, $y_{i}$ and $f(x_{i})$ are the measured and theoretically prediected values, respectively. The $R^{2}$--values for both $-$\SI{2.54}{\milli\meter} and $+$\SI{2.54}{\milli\meter} fiber shifts were calculated to be \SI{99.8}{\percent}. This excellent quantitative agreement validates our measurement apparatus as a collimation test system. With the obtained focal lengths, the $\Delta_{\text{PV}}=$\SI{9.53}{\micro\meter} is translated to the maximum divergence angle of $\theta_{\text{PV}}=$\SI{19.1}{\micro\radian}.

We further characterized the uniformity of our measurement precision over the range of scanning pentaprism. We recorded the centroid positions for five different fiber positions in the x-direction at each pentaprism position, as shown in Figure~\ref{fig:Result1}(c). The ratio of the centroid shift measured at the camera to the distance of the fiber translation is given by the focal ratio of the DUT and the imaging lens: $f_{2}/f_{1}$. We performed the least-squares regression with the linear function $y=a_{1}x+a_{2}$ to the measured spot positions and fiber positions at every pentaprism position. The obtained slope value $a_{1}\pm\delta{a}_{1}$ at each pentaprism position measures the focal ratio $f_{2}/f_{1}$, where the uncertainty of the regression $\delta{a}_{1}$ is given by the diagonal elements of a covarience matrix. Then, we verified the focal length of the imaging lens $f_{2}=\bar{a}_{1}/f_{1}=$\SI{500}{\milli\meter} with $\bar{a}_{1}$ denoting the averaged slope value over the scanning range. In our approach, the precision of estimating the fiber positions from the measured centroid positions at the camera is given by the uncertainty of imaging the fiber position $\delta\bar{a}_{1}\times f_{1}/f_{2}$. The averaged uncertainty was obtained to be $\delta\bar{a}_{1}\times f_{1}/f_{2}=0.009$, meaning that the centroid values can be determined with the uncertainty better than \SI{1}{\percent} over the entire scanning range. Note that $a_{1}\times f_{2}/f_{1}=-1$ (inversed image). The uniformity of the estimation over the scanning range is quantified by the variation of the obtained mean value $a_{1}\times f_{2}/f_{1}=-1\pm0.003$, as shown in Figure~\ref{fig:Result1}(d).

\begin{figure*}[t]
	\centering
	\includegraphics[width=0.7\linewidth]{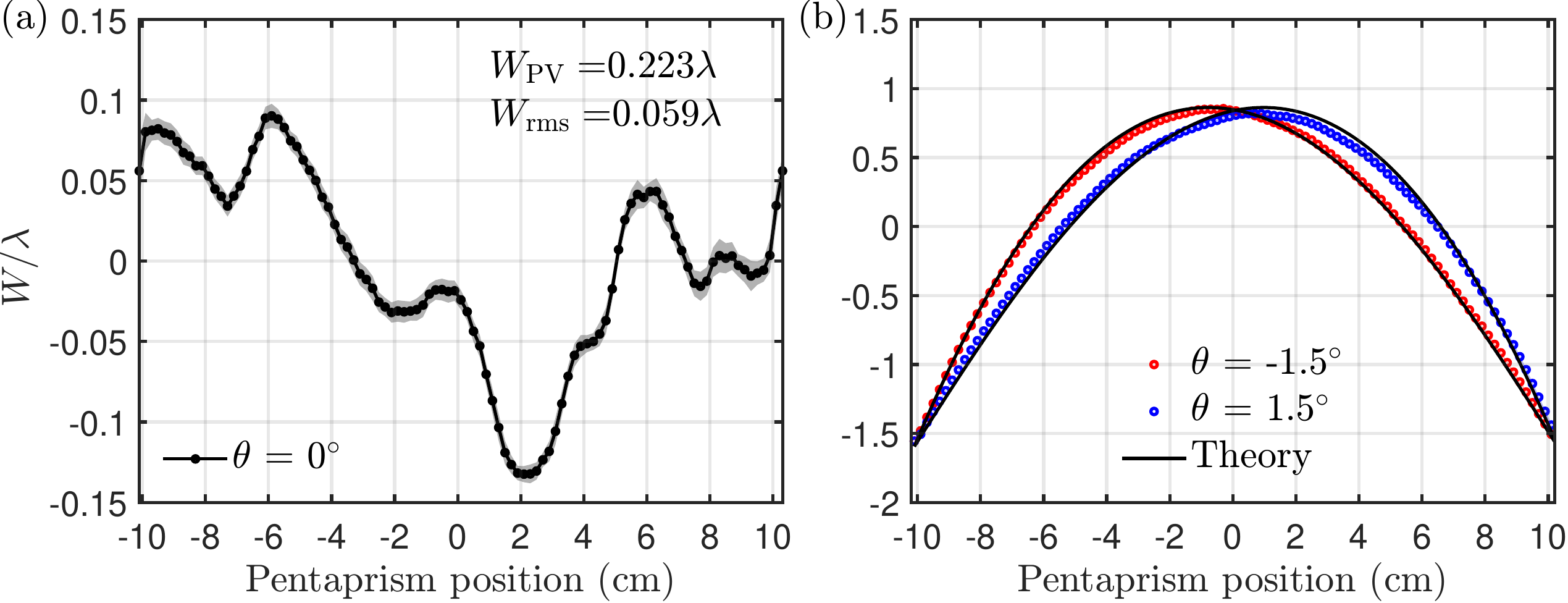}
	\caption{Reconstructed wavefront of the transmitted light from (a) the aligned lens and (b) tilted lens. Black solid lines represent the theoretical prediction calculated from a three-dimensional raytracing.
	}
	\label{fig:Result2}
\end{figure*}

\subsection{Wavefront reconstruction}

We reconstruct the transmitted wavefront from the measured centroid positions shown in Figure~\ref{fig:Result1}(a). The wavefront is normalized by the wavelength of $\lambda$=\SI{785}{\nano\meter}, as shown in Figure~\ref{fig:Result2}(a). The averaged uncertainty over the \SI{20}{\centi\meter} travel range of the pentaprism is estimated to be $\delta W=$0.007$\lambda$. This exceptional precision is comparable with the performance of shearing interferometers~\cite{Leibbrandt:96,Welsh:95}. Most previous reports on the wavefront measurement for large optics with Shack-Hartmann sensors exhibited the precision of order of 0.02$\lambda$ to 0.1$\lambda$~\cite{Neal2003,Forest2004,Kiikka2006}. This outperformed precision of our scheme is attributed to excellent stability of our scanning pentaprism system and the long focal length of the imaging lens.

We characterized aberrations for a tilted DUT and compared the result with theoretical prediction. The test optic was rotated by $\pm$\SI{1.5}{\degree} around the vertical y-axis. The measured wavefront at each tilt angle was subtracted from the wavefront measured at the normal angle, as shown in Figure~\ref{fig:Result2}(b), which removes the thickness variation of the DUT from manufacturing imperfections. We calculated the transmitted wavefront from our raytracing model (black solid lines in Figure~\ref{fig:Result2}(b)). The $R^{2}$ parameters were calculated to be \SI{99.9}{\percent} and \SI{99.6}{\percent} for the --\SI{1.5}{\degree} and +\SI{1.5}{\degree} tilt angles, respectively. The quantitative agreement between theory and experimental results further validates our wavefront measurement system with the scanning pentaprism.

\section{Conclusion}\label{sec:conclusion}

We developed a practical characterization system for optical aberrations of laser launching telescopes. The direct measurement of local wavefront slopes across output aperture of the test optic identifies the collimation of the transmitted light. We demonstrated transverse linear measurements of the transmitted wavefront and analyzed the measurement apparatus using a three-dimensional raytracing method. The test optic was chosen to be \SI{20.3}{\centi\meter} diameter achromatic doublet developed for the QEYSSat mission. With sufficiently redundant measurements and statistical analysis, it was shown that our wavefront-measurement system exhibits the precision better than 0.01$\lambda$. We compared the measured wavefronts with the theoretical predictions and excellent agreement between the two results validates our methodology.

Our scheme can be readily extended to two-dimensional measurement of wavefronts. The two-dimensional topography of optical surfaces has been obtained in scanning pentaprism deflectometries~\cite{Su2008,Allen2010,Qi2019}. One could perform additional linear measurements in different scanning directions, and reconstruct wavefronts using similar methods. For example, wavefronts can be expanded in terms of Zernike polynomials, and its gradient in the scanning directions can be fit to the measured slopes via a least-squares method, as discussed in~\cite{Qi2019}. Also, the measurement precision may be further improved by more precise alignment of the pentaprism. The quadratic response of centroid position shifts caused by the pentaprism $\alpha$ rotation (around the x-axis) can be minimized by the alignment of the pentaprism to the normal incident angle. This can be experimentally tested by centroid position measurements as a function of pentaprism rotation angles in both scanning and cross-scanning directions. The detailed description of the test method is provided in~\cite{Geckeler2006}. 

Our transverse linear measurement apparatus can be implemented at reasonable costs. The simple optical configuration requires minimal efforts for alignment, yet the system performs excellent stability and precision to characterize aberrations of large collimators. The wavelength range of our measurement setup is limited by the spectral sensitivity of the camera which can be switched without major modifications. We believe our practical test platform for characterizing aberrations is a useful tool for both scientific and industrial applications in diverse fields including optical satellite communications and astronomical observatories.

\section*{Acknowledgement}
We acknowledge Jean-Phillipe Bourgoin for his contributions to initial measurement. We acknowledge INO for suggesting the scanning pentaprism technique. Y.S.L acknowledges support from the Mike and Ophelia Lazaridis Fellowship Program. This research was supported in part by the Canadian Space Agency; Canada Foundation for Innovation (25403, 30833); Ontario Research Foundation (098, RE08-051); Canadian Institute for Advanced Research; Natural Sciences and Engineering Research Council of Canada (RGPIN-386329-2010); Industry Canada.

\section*{Author Declarations}
\subsection{Conflict of interest}
The authors have no conflicts to disclose.

\section*{Data availability}
The data that support the findings of this study are available from the corresponding author upon reasonable request.

\bibliography{references}

\end{document}